\documentclass{cernyrep}
\usepackage[T1]{fontenc}
\usepackage[bookmarks, colorlinks=true, linktoc=page, pdftex, linkcolor=black, citecolor=black, urlcolor=blue]{hyperref}
\usepackage{fancyhdr}
\fancyhfoffset{4 mm}
\fancypagestyle{ARTTITLE}{%
\fancyhf{} 
\lhead{\hfill CERN Accelerator School Proceedings ---
{\it RF for Accelerators} ---  Berlin, Germany,  2023\hfill}
\lfoot{\hspace{3mm} Available online at \url{https://cas.web.cern.ch/previous-schools}}
\rfoot{\thepage\hspace*{3mm}}
 
}

\usepackage{varwidth}
\usepackage{xcolor}
\begin{document}

\title{Transverse Deflecting Cavities}
\author{Graeme Burt}
\institute{Cockcroft Institute, Lancaster University, UK}

\begin{abstract}
Many applications in particle accelerators require a transverse deflection of the~beam that varies in time, to either rotate the beam or to provide a variable kick to different bunches in a train. One way of achieving this is to use an~RF cavity that deflects rather than accelerates the beam, with the sinusoidal RF voltage providing a time variation. These cavities require finite transverse electric and magnetic fields along the path of the beam through the cavity and hence different modes are required. This results in some variation in cavity geometry, beam-loading, and mode-damping which will be examined in this chapter.

\end{abstract}
\keywords{Transverse deflecting; crab.}
\maketitle
\thispagestyle{ARTTITLE}

\section{Introduction}
Transverse deflecting cavities are required in accelerators for a variety of applications that require a time varying transverse deflection of charged particles. These are RF cavities where the Lorentz Force acting on a charged particle is transverse to the direction of motion, acting to deflect the beam at a small angle relative to the central axis, which we will refer to as a kick. The kick can be phased with the peak of the~sinusoidal varying transverse voltage to deflect the entire bunch, or at the zero crossing such that the head and tail receive equal and opposite kicks, causing an effect similar to a rotation for small kicks known as a crabbing. These can be used to separate different bunches to different experiments, to rotate the bunch for diagnostic or alignment purposes. 

\begin{itemize}
  \item Particle separation (Kaon separated from pions) where two deflecting cavities are separated in distance and phases so the time of flight of one particle type causes the two deflections to cancel but not of other species.
  \item Temporal beam diagnostics, where the kick is dependant on longitudinal position such that the~longitudinal profile can be projected on to a screen
  \item Crab-crossing in colliders, where the crossing angle at the collision point causes an angle between the beams at the interaction point this can be corrected with a transverse deflecting cavity known as a crab cavity.
  \item Emittance exchange where a deflecting cavity placed inside a magnetic chicane causes the longitudinal and transverse phase space to swap.
  \item Beam splitters (to different user stations) where the beam separation is not an integer multiple of the RF period every bunch gets a different deflection.
\end{itemize}

\section{Transverse Kicks and Rotation}
If we apply a kick of voltage $V_t$ to a charged particle it gain a transverse energy $qV_t$

\begin{figure}[!b]
\centerline{\fbox{\hbox to 35pc{\vbox to 8pc{
\includegraphics[width=0.9\linewidth]{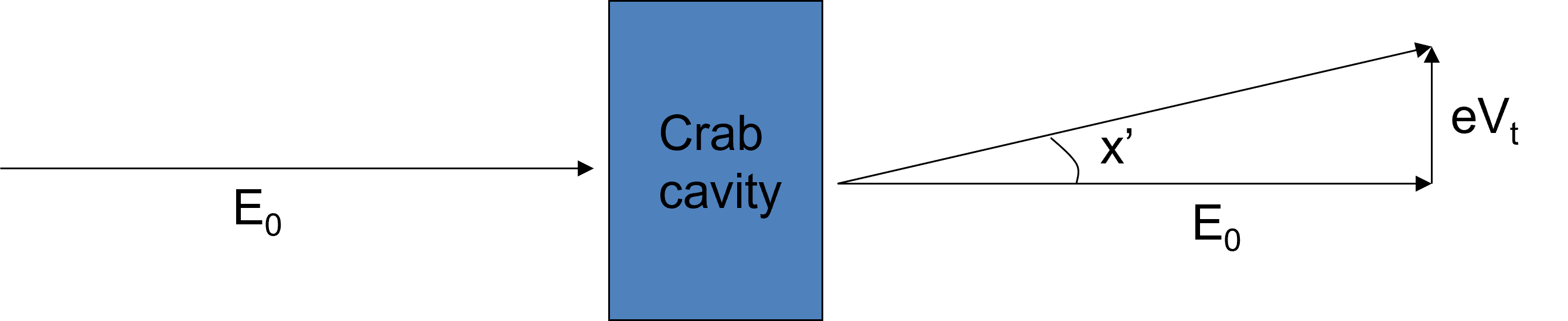}}}}
}
\caption{\label{fig:Kick}The transverse divergence, x', caused by a transverse voltage, $V_{t}$, applied to a beam of energy $E_{0}$.}
\end{figure}

If the particle has a longitudinal energy $E_0$ then the electron will have a trajectory with angle, $x' = arctan(\frac{qV_t}{E_0}) \approx \frac{qV_t}{E_0}$, as shown in Fig. \ref{fig:Kick}.

The transverse offset of a bunch that has been kicked by a transverse deflecting cavity is dependant on the beam optics between the cavity and the point where we measure the offset. The relationship between a particles transverse position and velocity between two positions is given by the $R_{12}$ element of the transfer matrix, R. A particle with a transverse velocity of zero before the deflecting cavity which leaves the cavity at an angle, x’, will have an offset at a later point, $\Delta x$, given by $\Delta x  = R_{12} \frac{qV_t}{E_0}$.

Transverse deflecting cavities can also be used to rotate bunches as well as deflect them as they can deliver a sinusoidal time-varying transverse voltage. If we consider a number of charged particles that enter a deflecting cavity at a time t, the offset of these particles at a position further down the accelerator is given by $\Delta x  = R_{12} \frac{qV_tsin(\omega t)}{E_0},$ where $\omega$ is the resonant frequency of the cavity. Hence a particle entering the cavity at time t=0 will not experience a transverse kick while particles leaving the cavity just before and just after this will experience equal and opposite kicks. As the bunch travels down the accelerator the~transverse momentum will cause the head and tail to move away from the centre axis, hence the~bunch will appear to have rotated about the particle which left the cavity at t=0. If the bunch duration, $\tau$, is small compared to the~period of the RF then the rotation, $\theta$, is given by $\theta = \frac{\Delta x}{c \tau} = R_{12} \frac{qV_tsin(\omega \tau)}{c \tau E_0}$  which for small angles can be approximated as
$\theta \approx R_{12} \frac{qV_t \omega}{c E_0}$. Deflecting mode cavities phased, $\omega t = 0$, to give bunch rotations rather than deflections are known as crab cavities.

To provide a transverse force to charged particles we must provide transverse electric fields, E, transverse magnetic fields, B, or both as given by the Lorentz force equation, $F=q(E + v \times B), $
where q is the particles charge and v is the particle velocity. As the particles travel in the z direction, for a~horizontal force we require a horizontal electric field and/or a vertical magnetic field. The electric field provides an additional traverse momentum, while a magnetic field changes the beams path providing transverse momentum at the expense of longitudinal momentum. If the beam has a high enough kinetic energy such that the additional transverse momentum provided by a transverse electric field is very small compared to the longitudinal momentum, then the effect of electric and magnetic fields are virtually the~same. In this case it is standard to integrate and combine both fields into a single transverse voltage, $V_t$,  
$V_t= \int{(E_t + v_z \times B_t)}{dz},$
where $v_z$ is the longitudinal component of the beam velocity and $E_t$ and $B_t$ are the transverse electric and magnetic field components. Note that this is not a real voltage as some of the force comes from the magnetic field, it is really transverse work per unit charge. The electric and magnetic fields will be out of phase for a standing wave cavity, however due to the transit time of the~particles and the~different mode patterns the integrated E and B fields may have any phase between them. For a fully relativistic beam and a cavity which is a half wavelength long the E and B fields will either be exactly in phase, acting together, or exactly out of phase, cancelling each other.

For a pillbox cavity the transverse fields of mode $TM_{mnp}$ vary transversely as $J_m(k_tr)sin(m \phi)$ for $B_r$ and $E_{\phi}$ and $J’_m(k_tr)cos(m \phi)$ for $B_{\phi}$ and $E_r$, where $k_t$ is the transverse wavenumber. Also the transverse fields of mode $TE_{mnp}$ vary transversely as $J’_m(k_tr)cos(m \phi)$  for $B_r$ and $E_\phi$ and $J_m(k_tr)sin(m \phi)$  for $B_\phi$ and $E_r$. The only Bessel functions that are finite at r=0 are $J_0(0)$ and $J'_1(0)$, hence as the m=0 TM modes do not have $B_r$ and $E_{\phi}$ components only modes of order m=1, the dipole modes, have non-zero transverse fields on the central axis at r=0. Monopole modes such as the accelerating $TM_{010}$ mode of a pillbox cavity only have longitudinal fields on axis so they cannot be used to deflect the beam if the beam travels down the centre, although do have transverse fields off-axis. Most dipole modes have both transverse electric and transverse magnetic fields at r=0 however these two fields can either add constructively or cancel each other out. In order to find the transverse kick for ultra-relativistic particles we instead use Panofsky-Wenzel theorem \cite{bib:Panofsky} which relates the transverse voltage, $V_t$, to the~transverse variation of the longitudinal electric field.

\begin{equation}
V_t= - \frac{ic}{\omega} \int{\nabla_t E_z}{dz},
\end{equation}
where $\nabla_t$ is the transverse components of the del operator. This equation is valid for all cavities, where the~beam voltage is v=c, and is not constrained to pillbox cavities.  As the transverse voltage is proportional to the gradient of $E_z$ it is clear that TE modes, which do not have longitudinal electric fields, cannot be used to provide a transverse deflection of ultra-relativistic particles. To deflect ultra-relativistic particles using a pillbox cavity we must use TM dipole modes. TE modes can be used to deflect low velocity particles where the Panofsky-Wenzel theorem does not apply.

\begin{figure}[!h]
\centerline{\fbox{\hbox to 17pc{\vbox to 22pc{
\includegraphics[width=0.45\linewidth]{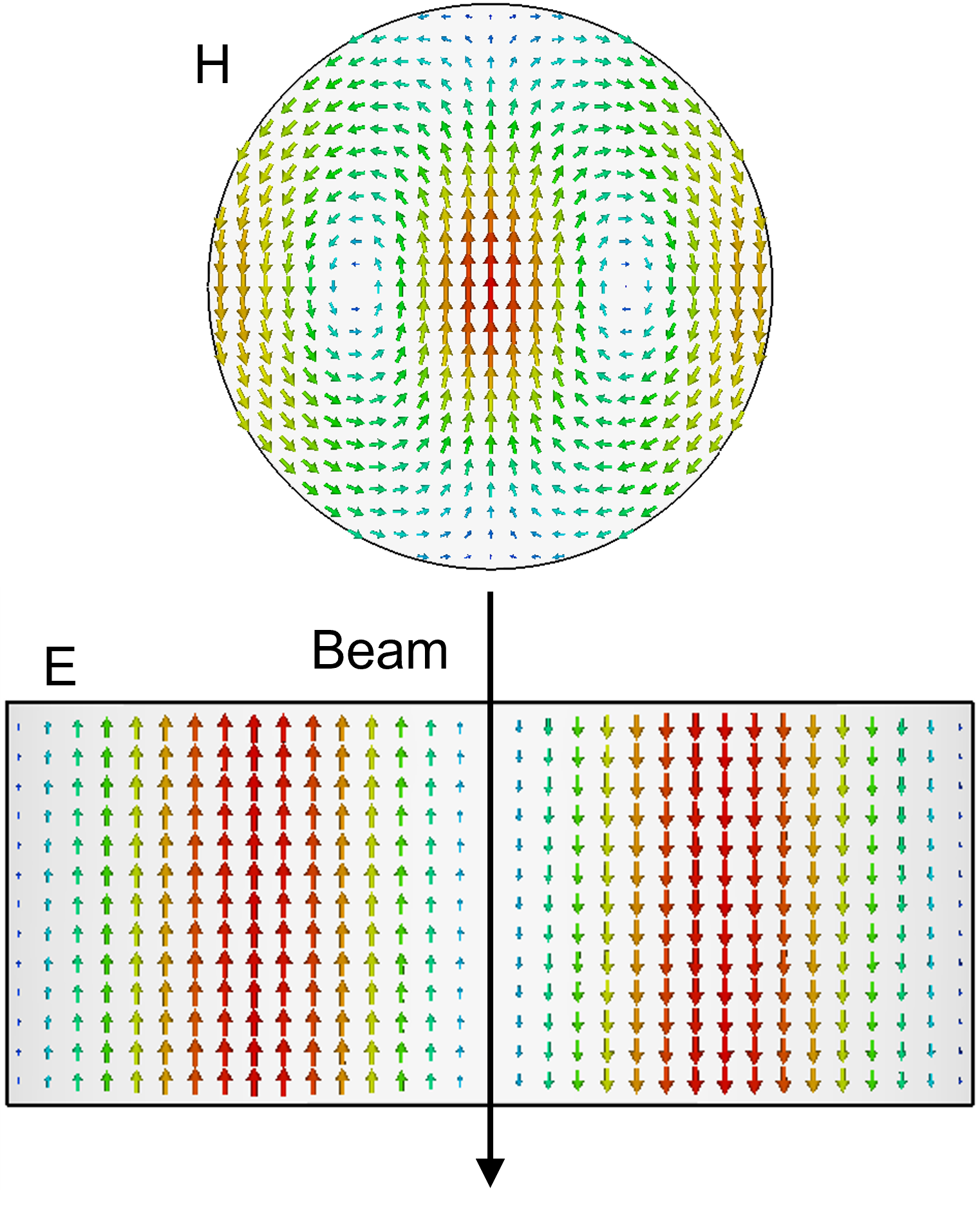}}}}
}
\caption{\label{fig:TM110} The fields of the TM110 mode in a pillbox cavity.}
\end{figure}

The lowest frequency dipole modes in pillbox cavities are usually the $TE_{111}$ mode and the $TM_{110}$ mode. The $TE_{111}$ mode has both transverse electric and magnetic fields on axis, perpendicular to each other, however as already mentioned they both exactly cancel each other out as they are 180 degree out of phase with each other. The $TM_{110}$ mode has zero electric field on axis but has a large transverse magnetic field which is used to provide the transverse force, as shown in Fig. \ref{fig:TM110}. In a pillbox cavity the~dipole fields of a $TM_{010}$ mode are given by 

\begin{align}
E_z &= E_0 J_1 \left(r \frac{\zeta_{01}}{a} \right) \cos \left(\phi\right)  \exp \left(i \omega t\right),\nonumber	\\
E_r &= 0,\nonumber	\\
E_{\phi} &=0,\nonumber	\\
H_r &= 0,\nonumber	\\
H_{\phi} &= \frac{i \omega \epsilon}{k_t} E_0 J_1' \left(r \frac{\zeta_{01}}{a} \right) \cos \left(\phi\right) \exp \left(i \omega t\right) ,\nonumber	\\
H_z &= 0,	
\end{align}
where $J_m'(x)=\mathrm{d}J_m(x)/\mathrm{d} x$.

\begin{figure}[!b]
\centerline{\fbox{\hbox to 30pc{\vbox to 20pc{
\includegraphics[width=0.9\linewidth]{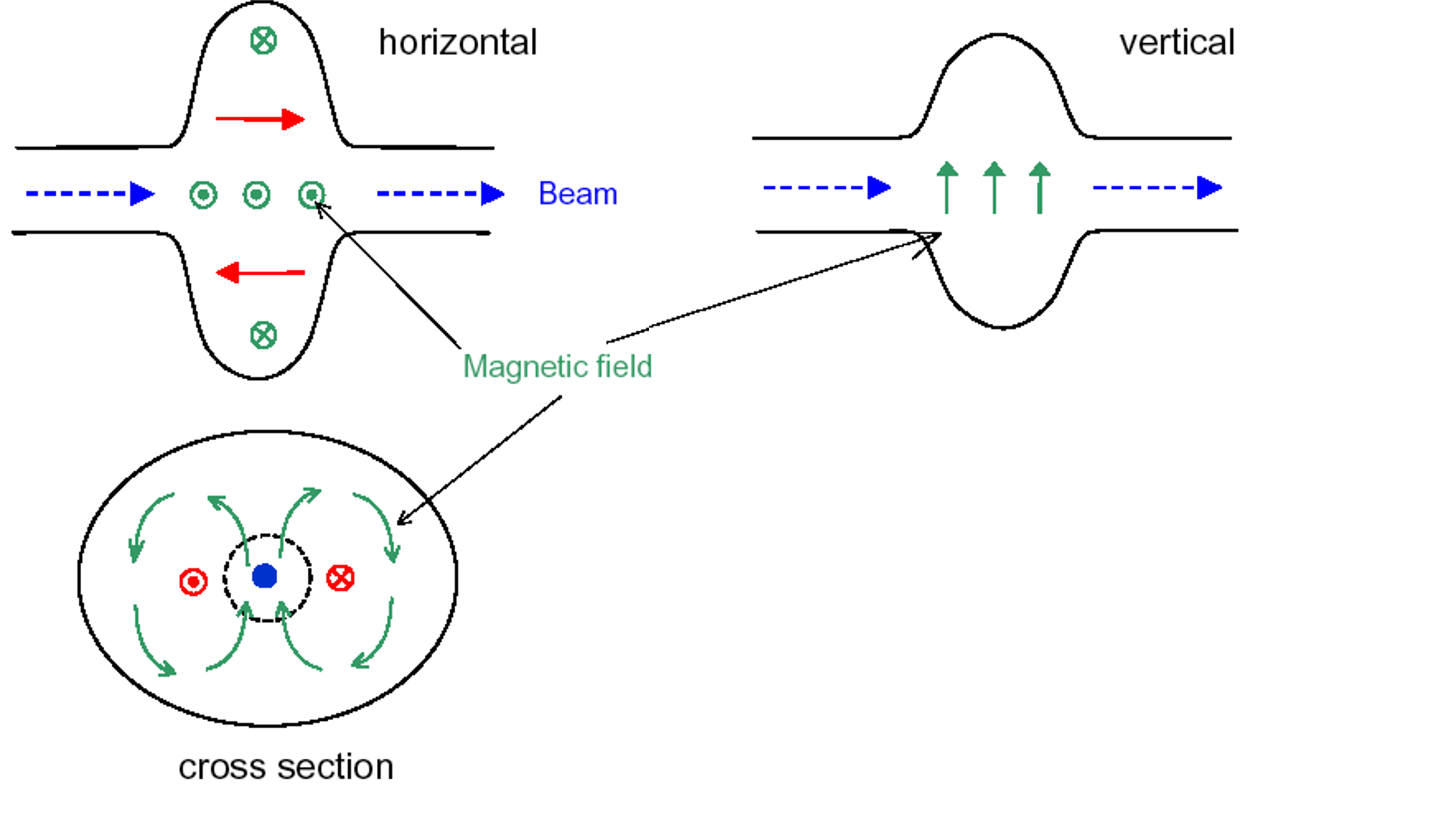}}}}
}
\caption{\label{fig:single} The fields inside an elliptical crab cavity with beampipes.}
\end{figure}

The addition of the beam-pipes and iris’ causes the $TM_{010}$ mode in the cavity to couple to a~$TE_{110}$-like mode in the iris causing a transverse electric field to appear on axis, making the mode a~hybrid mode, shown in Fig. \ref{fig:single}. This interaction has a marked effect on the dispersion curves of the~$TM_{010}$-like and $TE_{111}$-like modes of the cavity, which we discuss later. An electron travelling through the cavity at close to the speed of light will experience the electric fields in the iris and the magnetic fields in the~cavity, however as the cell periodicity is usually half the wavelength and the electric and magnetic fields are 180 degrees out-of-phase in standing wave cavities the forces due to the two fields act in the~same direction for deflecting mode cavities. The fields near the cavities axis of longitudinal symmetry are given by,

\begin{align}
E_z &= E_0 x \cos (k z) \cos(\omega t),\nonumber	\\
E_x &= E_0 \frac{k}{4} (a^2+x^2-y^2) \sin (k z) \cos(\omega t),\nonumber	\\
E_y &=E_0 \frac{k}{2} x y \sin (k z) \cos(\omega t),\nonumber	\\
c B_z &= E_0  y \sin (k z) \sin (\omega t),\nonumber	\\
c B_x &= E_0 \frac{k}{2} x y \cos (k z) \sin (\omega t),\nonumber	\\
c B_y &= E_0 \frac{1}{k} \left( \frac{(k a)^2}{4}-1+ \frac{k^2(x^2-y^2)}{4} \right) \cos (k z) \sin (\omega t),
\end{align}

For the TM110-like dipole modes we can insert the equation above into the Panofsky-Wenzel equation to obtain the relationship between $V_x$ and $V_z(x)$ measured at some distance off axis, x, 

\begin{align}
V_t &= \frac{-ic}{\omega x} V_z(x).
\end{align}

For a azimuthally symmetric cavity $V_z$ is zero at the cavity centre. The definition of a transverse voltage leads to the definition of a transverse shunt impedance $R_t$ which is useful for relating the transverse voltage $V_x$ to power losses and longitudinal voltage $V_z$ at an off-axis distance of a in the cavity.

\begin{align}
R_t &= \frac{V_t^2}{2P}=\frac{c^2}{(\omega x)^2} \frac{V_z(x)^2}{2P}.
\end{align}

\newpage
\section{Applications of Transverse deflecting cavities}
\subsection{RF Separators}
The first transverse deflecting cavities were used as RF separators, which were able to separate particles of different masses \cite{bib:RFseperator}. Each particle traversing the cavity receives an additional transverse momentum kick, but particles of different masses, such as kaons and protons, will be deflected at different angles. This leads to a transverse separation between the two particle species which can be used to separate them for detection. One such system proposed at Fermilab utilises two deflecting cavities separated by a lens system \cite{bib:Kaon}. The first cavity provides a transverse kick to all particles which then travel through the lens system, as the pions, kaons and protons will all arrive at the second cavity at different times they will all receive differing kicks. The pions and protons are deflected into a beam stopper and the kaons are deflected to a detector. For this scheme the phase separation between the pions and kaons, $\Delta \phi_k$

\begin{align}
\Delta \phi_k  = \frac{\omega L}{c} \Big( \frac{1}{\beta_k}-\frac{1}{\beta_\pi} \Big),
\end{align}
where $\omega$ is the angular frequency of the RF cavity, L is the distance between the two Rf systems, c is the~speed of light and $\beta_k$/$\beta_\pi$ are the normalised velocities of the kaons and pions.

Another application of transverse separators is to separate trains of electrons to send them to different detectors, insertion devices or for use in Energy Recovery Linacs’s \cite{bib:ERLseperator}. In such a system the~RF frequency of the deflecting cavity is a sub-harmonic of the bunch separation hence several bunches traverse the cavity in one RF period. Each of these bunches will arrive at a different phase and hence will receive a different transverse kick, allowing bunches to be separated.

\subsection{RF cavity based bunch length diagnostics}
If we alter the phase of a transverse deflecting cavity, instead of deflecting the whole bunch, we can introduce a correlation between a particles transverse momenta and its longitudinal position in the bunch, by having the bunch phased with the linear part of the sinusoidal oscillation in the transverse voltage.
\begin{figure}[!b]
\includegraphics[width=0.9\linewidth]{Images/Rotation.png}
\centering
\caption{\label{fig:Rotation}The rotation applied to a bunch centred on the zero crossing phase of a deflecting cavity.}
\end{figure}

This property can be exploited for use in a bunch length diagnostic measurement where the bunch is directed onto a screen and the bunch length can hence be calculated by the voltage gradient across the bunch and the transverse size of the bunch on the screen. The bunch will rotate as it travels to wards the screen, shown in Fig. \ref{fig:Rotation}, such that the transverse position of the screen is correlated to the~longitudinal position in the bunch. One must be careful that the transverse deflection causes the bunch to be sufficiently spread out transversely at the screen such that the transverse bunch size doesn’t dominate the~measurement. The temporal resolution, $\Delta t$ of such a measurement is given by

\begin{align}
\Delta t  = \sigma_x  \frac{2 m c^2 \gamma}{\omega L e V_t},
\end{align}
where $\sigma x$ is the bunch width, $\gamma$ is the relativistic correction facor, L is the distance between the cavity and the screen, e is the charge of the electron, and $V_t$ is the transverse voltage \cite{bib:diagnostic}.

\subsection{Crab Cavities for Colliders}
\begin{figure}[!b]
\centerline{\fbox{\hbox to 35pc{\vbox to 13pc{
\includegraphics[width=0.9\linewidth]{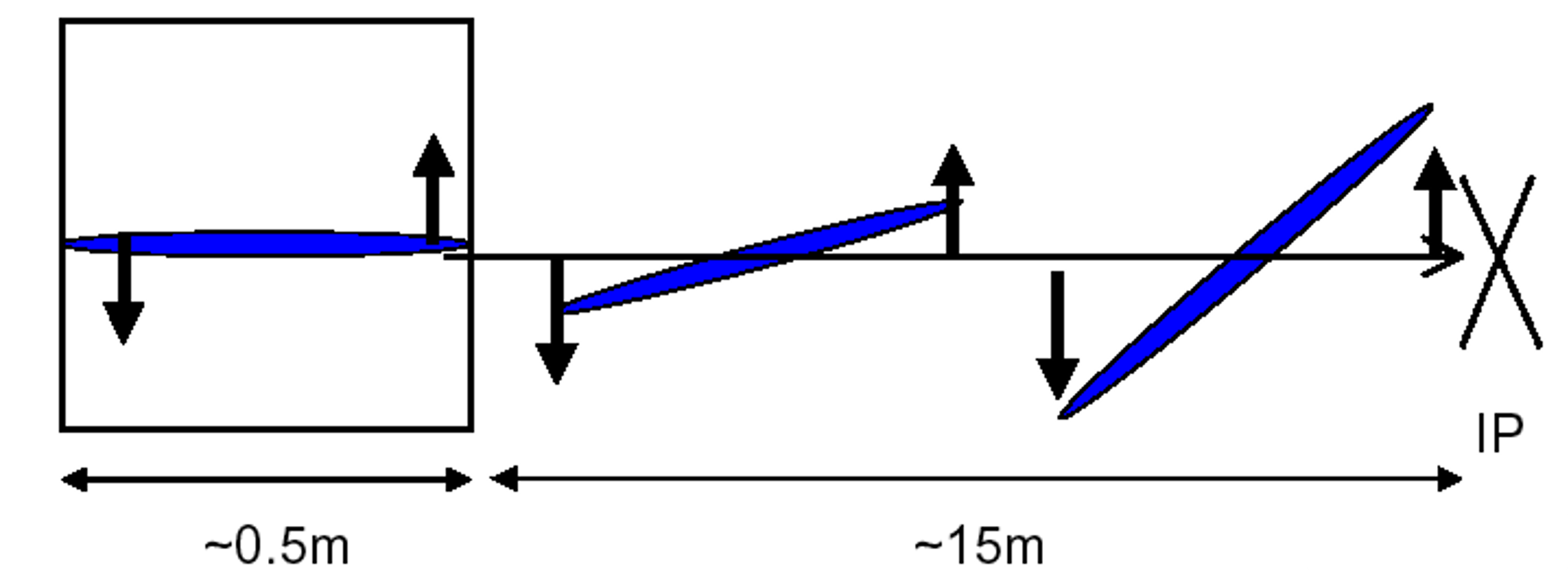}}}}
}
\caption{\label{fig:Crabbing} The crabbing of a bunch to rotate its alignment while travelling towards the interaction point.}
\end{figure}

In circular colliders such as KEK-B, and LHC the colliding beams circulate in two separate beam-lines, at the interaction region (IR) the two beams are brought together with a finite crossing angle and then separated again. Also the next generation of linear electron-positron colliders (ILC and CLIC) are proposed to utilise a finite crossing angles at the interaction region (IR) to simplify extraction of the spent beam. However these finite crossing angles reduce the luminosity of the collision as the two beam cross sections do not completely overlap. Simple geometric arguments can be used to show that the luminosity reduction factor, S, can be given as

\begin{align}
S  =  \frac{1}{\sqrt{1+\big(\frac{\sigma_c \theta_c}{2 \sigma_x}\big)^2}},
\end{align}
where $\theta_{c}$ is the full crossing angle, $\sigma_{x}$ is the bunch width and $\sigma_{z}$ is the bunch width \cite{bib:crabcavities}. In order to correct for the finite crossing angle it is necessary to rotate the bunch prior to the IR. One method of rotating the bunches, suggested by R. Palmer \cite{bib:Palmer}, is to use crab cavities. Crab cavities are a subset of transverse deflecting cavities, where the bunch centre traverses the cavity at the zero kick phase, similar to a bunch length diagnostic deflector. Due to the finite size of the bunch only the centre does not receive a kick, the~head and tail of the bunch receive equal and opposite kicks, shown in Fig. \ref{fig:Crabbing}. 

For linear colliders the required voltage can be calculated by considering the required offset of an~electron at the head of a bunch at the IR, $\Delta x_{h}$, in order for the beam to have the correct rotation. From this offset we can use the $R_{12}$ transfer matrix component between the crab cavity location and the IR to estimate the required transverse voltage, $V_{t}$.

\begin{align}
\Delta x_{h}  = R_{12} \frac{qV_t sin(\frac{\omega \sigma_{z}}{c})}{E_0},
\end{align}
where $E_{0}$ is the beam energy, $\omega$ is the frequency of the cavity and c is the speed of light. As can be seen increasing the $R_{12}$ matrix component, which relates x’ at the cavity to $\Delta x$ at the IR, reduces the required voltage, hence crab cavities tend to be placed at locations with a high $R_{12}$. This can lead to problems as instabilities at the cavity, such as RF phase errors and wakefields, can be magnified at the IR.

For circular colliders there are two possible scenarios for crab cavities, a local scheme and a global scheme. In a local scheme the beam is rotated by a crab cavity close to the IR and straightened by a 2nd crab cavity place on the other side of the IR, each at a phase advance of $\psi$/2 from the IR. For this case the voltage required from the first cavity, $V_1$, is

\begin{align}
V_1  = \frac{c^2 p_s tan(\frac{\theta_c}{2})}{q \omega \sqrt{\beta* \beta_{crab}} sin(\Delta \phi_0)},
\end{align}
where $p_s$ is the particle momentum, q is the particle charge, $\beta *$ is the beta function at the IR, $\beta_{crab}$ is the beta function at the crab cavity, and $\Delta \phi_0$ is the phase advance between the cavity and the IR [10]. The~voltage required from the second cavity, $V_2$, is

\begin{align}
V_2  = R_{22} V_1.
\end{align}

Alternatively a global scheme can be utilised where a single crab cavity is used per beam-line to rotate the bunch. As the bunch trajectory is a closed orbit the bunch will rotate as it travels round the ring achieving the correct rotation at the IR. For this scheme the required voltage is

\begin{align}
V  = \frac{c^2 p_s tan(\frac{\theta_c}{2})}{q \omega \sqrt{\beta* \beta_{crab}} sin(\Delta \phi_0)} \Big|\frac{2 sin(\pi Q)}{cos(\Delta \phi_0 - \pi Q} \Big|,
\end{align}
where Q is the betatron tune of the storage ring \cite{bib:globalcrab}. 
For both circular and linear crab cavities the cavity roll alignment is critical. If the cavity is misaligned the crab cavity will create an additional crossing angle in the vertical plane reducing luminosity. For beams that are focussed down to nanometre scale vertical beam sizes at the IP the roll alignment tolerances can be very tight.

\section{Dispersion and Equivalent Circuit of Dipole Modes}
\begin{figure}[!b]
\centerline{\fbox{\hbox to 35pc{\vbox to 16pc{
\includegraphics[width=0.9\linewidth]{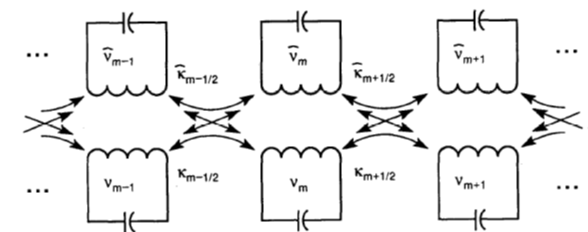}}}}
}
\caption{\label{fig:Twochain}The two chain equivalent circuit model for multicell dipole mode cavities \cite{bib:Bane}.}
\end{figure}

To analyse the band structure of most multi-cell cavities an equivalent circuit is traditionally used. For accelerating cavities a single chain of LC resonators which are capacitively coupled can be used to accurately predict the modal frequencies and cell amplitudes. For dipole modes the crabbing mode is made up from a mixture of TE and TM modes hence a single chain or resonators is not sufficient to model such a cavity. For dipole modes a better fit can be made by utilising a double chain model of LC resonators \cite{bib:Bane}, one for the TE mode and one for the TM mode, coupled inductively (where a negative inductance can represent a capacitive coupling) as shown in Fig. \ref{fig:Twochain}. This circuit has two solutions giving a TM-like hybrid pass-band and a TE-like hybrid pass-band.

\begin{figure}[!b]
\centerline{\fbox{\hbox to 35pc{\vbox to 21pc{
\includegraphics[width=0.9\linewidth]{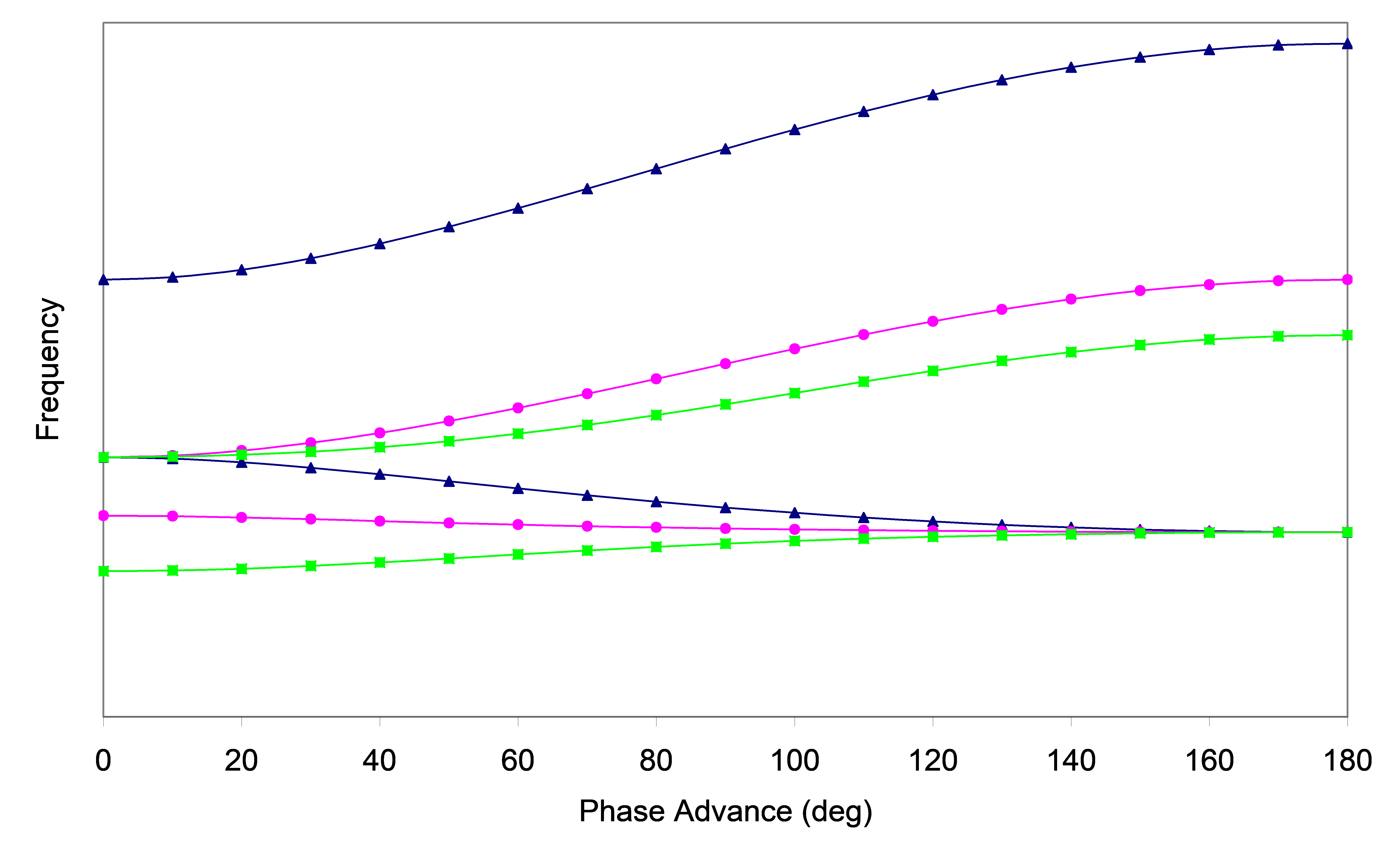}}}}
}
\caption{\label{fig:Dipolepassband} The two passbands of a dipole mode in a multicell cavity for three different iris radius.}
\end{figure}

The circuit equation for the $m^{th}$ cell of this circuit is

\begin{align}
\Big( \frac{1}{v^2}-\lambda \Big) f_m -\frac{\kappa}{2} f_{m+1} -\frac{\kappa}{2} f_{m-1} = - \frac{\sqrt{\kappa \bar{\kappa}}}{2} f_{m+1} - \frac{\sqrt{\kappa \bar{\kappa}}}{2} f_{m-1} \\
\Big( \frac{1}{\bar{v}^2}-\lambda \Big) \bar{f_m} -\frac{\bar{\kappa}}{2} \bar{f_{m+1}} -\frac{\bar{\kappa}}{2} \bar{f_{m-1}} = \frac{\sqrt{\kappa \bar{\kappa}}}{2} \bar{f_{m+1}} - \frac{\sqrt{\kappa \bar{\kappa}}}{2} \bar{f_{m-1}},
\end{align}
where f and $\bar{f}$  are the amplitude variables, $\lambda$  is the eigenvalue, v  and $\bar{v}$  are the frequencies, $\kappa$  and $\bar{\kappa}$  are the intra-chain coupling factors, and  $\sqrt{\kappa \bar{\kappa}}$ is the inter-chain coupling factor. The terms on the LHS are the terms for a chain of resonators while the RHS contains the cross coupling terms. The values for the~variables in these equations can be found by measuring or simulating the eigen-frequencies of the~$\pi$ and 0 modes of both pass-bands. 
For the uncoupled pass-bands, for example in a cavity with a~small iris, the~TM modes are magnetically coupled and the TE modes are electrically coupled, however the~intra-chain coupling is small because the iris is small.  As the iris size increases the intra-chain coupling increases and the TE pass-band drops in frequency and the cross coupling increases. When the~two pass-bands are close in frequency the two pass-bands do not cross over instead the TE pass-band pushes the TM modes with low phase advance downwards in frequency, known as an avoided crossing. This initially reduces the coupling between cells for the lower eigenmode and hence the group velocity and mode separation, as can be seen in the blue curve in Fig. \ref{fig:Dipolepassband}. As the iris size increases further the~eigenmode coupling starts to increase again, but with the opposite sign, as the lower band becomes more TE-like as can be seen in the green curve in Fig. \ref{fig:Dipolepassband}. This can be problematic as most accelerators require a~fairly large cell-to-cell coupling to increase the separation between modes. Moving to larger beam-pipes can increase the~coupling however this often results in high surface fields.

\section{Beamloading in transverse deflecting cavities}
Beam-loading is a net transfer of energy between the RF stored energy and the beam. As the current is directed in the longitudinal direction, transverse voltages do not lead to a transfer of energy, except a small component due to the transverse velocity of the beam. In an ideal accelerator there is no beam-loading in dipole cavities as the longitudinal electric field on axis is zero. However, in real accelerators particle beams rarely travel precisely on-axis and there is usually some offset from the ideal path. The~longitudinal electric fields in dipole modes vary linearly with displacement in the plane of deflection for small radial offsets. The power deposited or removed from the cavity from a train of bunches of current, $I_{t}$, traversing the cavity with a bunch offset in the deflecting plane, x, at a phase $\phi$ can then be given as \cite{bib:beamload}

\begin{equation}
P= V_{z} I_{t} sin(\phi) = V_{t} \frac{x \omega}{c} I_{t} sin(\phi),
\end{equation}
where $\phi$=0 is in phase with the deflecting voltage. It should also be noted that the transverse kick is 90 degrees out-of-phase with the longitudinal voltage and hence the beam loading, hence in a~deflecting phase there is no beam-loading and in crabbing phase the beam-loading is maximal. As the~longitudinal electric field can be either positive or negative depending on which direction the bunch is offset, the~beam-loading can either deposit energy in the cavity or take energy out.  The beam-loading can vary greatly between bunches and if the beam current is sufficiently large could in theory produce enough power to drive the cavity without an external power (although such a cavity would be unstable). This can cause problems as the beam offset can vary in time causing large beam-loading transients. In an SRF system where beam-loading is large this can be problematic as it means that the power coupler on a crab cavity cannot be matched to the cavity when beam is present as the beam loading cannot be predicted and the~amplifier/LLRF system must cope with large power transients.

\section{End effects at low energy}
In all cavities the decay of the cavity fields into the beampipe causes some of the field to decelerate the bunch as the phase change in time flips the field. This effect has only a minor effect in accelerating cavities; however, the effect can be more pronounced in deflecting cavities designed to rotate the beam. In a deflecting cavity phased such that the centeroid experiences no net kick, the bunch centeroid actually sees equal and opposite kicks as it traverses each cell, such that the kicks cancel over the cell. However, in the end cells the kick at the front and back of the cell is no longer symmetric, and the bunch centeroid will experience a net kick. Net kick to the bunch centeriod from the end cells at the start and end of the cavity however do cancel with each other so over a full cavity there is no net transverse momentum imparted to the centeriod. For high energy beams this in itself is not a problem as the kicks do indeed cancel, but at lower beam energies, when the beam is less stiff, the beam can drift transversely in the gap between the start and end of the cavity, and hence the beam will drift off-axis even though the transverse momentum is cancelled. This drift often requires a pair of corrector magnets at the entrance and exit of the cavity. Alternatively the end cell geometry can be optimised by reducing the length of the end cells to balance the kicks at the front and back of the cell \cite{bib:CLARAdeflector}.

\section{Peak surface fields in deflecting mode cavities}
Typically dipole mode cavities have a large surface magnetic field as the surface current flows around the~aperture. The surface electric field tends to be comparatively small as the charge maxima is away from the~aperture. Typically the ratio of peak surface magnetic field to peak surface magnetic field in an elliptical dipole cavity is around 4, which is twice the value expected in accelerating modes. In SRF elliptical deflectors the main limitation will hence be the magnetic field and lower field emission can be expected. It should be noted that compact TEM and TE mode crab cavities have much higher electric fields as we shall discuss later.
In normal-conducting cavities this adds some uncertainty on the maximum operating field of a dipole mode cavity. Traditionally the field limits are stated only in terms of peak electric field, and this would suggest that dipole modes could run at very high gradients. However the~latest research into breakdown in RF cavities suggest the magnetic field also plays a role. When high gradient tests have been applied to dipole mode cavities the results have been complex. The breakdown fields tend to be at a fairly high magnetic field and low electric field compared to accelerating cavity field, but at fields consistent with the modified Poynting vector \cite{bib:CLICcrab}. Interestingly in inspecting the breakdown damage it lines up perfectly with electric field, and is not consistent with the modified Poynting vector.

\section{Mode Polarisation}
The transverse kick delivered to the beam from a transverse deflecting cavity has a given polarisation, such that the direction of the kick is well defined. However the $TM_{110}$ in a pillbox is degenerate and has two modes with different polarisations, one in the horizontal plane and one in the vertical plane. A~transverse deflecting cavity will utilise one of these modes for its operation. The mode in the opposite polarisation to the operating mode is often referred to as the same order mode (SOM) \cite{bib:SOM}. As these modes are degenerate they have the same frequency and shunt impedance which can be problematic as the SOM will hence be resonant with the beam and will couple strongly to it. If the modes are at the same frequency then the SOM will be cut-off in the beam-pipe like the operating mode making it difficult to damp, in addition if we use cut-off waveguide or resonant filters to avoid dampers coupling to the operating mode we will also avoid coupling to the SOM as well. In order to separate these two modes we will polarise the cavity in some way, such as using an elliptical cross-section, coupling holes in the iris or using polarisation rods in the cavity By polarising the cavity we can shift the SOM to a higher frequency which will move it out of resonance with the beam and make it easier to damp. However making the cavity strongly asymmetric will reduce the performance of the operating mode so often dampers are placed in the plane of the SOM so that they do not couple to the operating mode as it has no fields in that plane. If the dampers do not couple to the operating mode the dampers can avoid using resonant filters or cut-off waveguide and the SOM doesn’t need as large a frequency separation for strong damping, This method has an inherent risk associated with it in that if there are mechanical errors in the cavity manufacture the mode polarisations can rotate slightly causing the operating mode to be damped.

\section{Lower Order Modes}

In a pillbox cavity with length equal to around half a wavelength, as is used in many deflecting mode cavities, the $TM_{110}$ mode is not the first resonant mode in the cavity.  The first resonant mode is the~$TM_{010}$ monopole mode which is commonly used in accelerating cavities. As this mode will be at a lower frequency than the operating mode we refer to this cavity as the lower order mode (LOM) and this mode will likely be trapped in the cavity as its frequency will be much lower than the beam-pipe cut-off frequency. This requires the use of special dampers to remove this mode from the cavity. In the KEK-B crab cavity a hollow coaxial beam-pipe couple was utilised where the TEM mode of the coax coupled to the LOM but due to symmetry did not couple to dipole modes \cite{bib:KEKBcrab}, however a notch filter was also used in case of misalignment. For the ILC crab cavity a hook type coaxial coupler was used, positioned at the field null of the deflecting mode to avoid damping it. This is similar to the standard F-probe type coaxial coupler used in superconducting cavities except the end is bent into a hook such that the long tip is bend towards the cavity.

\section{TEM and TE type deflecting cavities}
When we require compact RF cavities it is standard to use TEM mode coaxial cavities such as quarter-wave or half-wave resonators. Coaxial cavities are not ideal for deflecting mode cavities as they do not have the correct fields for deflection, however other types of cavities can support TEM modes or TE modes that have field patterns better suited to deflection. In order two obtain a deflection we have two options.

\begin{figure}[!h]
\centerline{\fbox{\hbox to 22.5pc{\vbox to 35pc{
\includegraphics[width=0.60\linewidth]{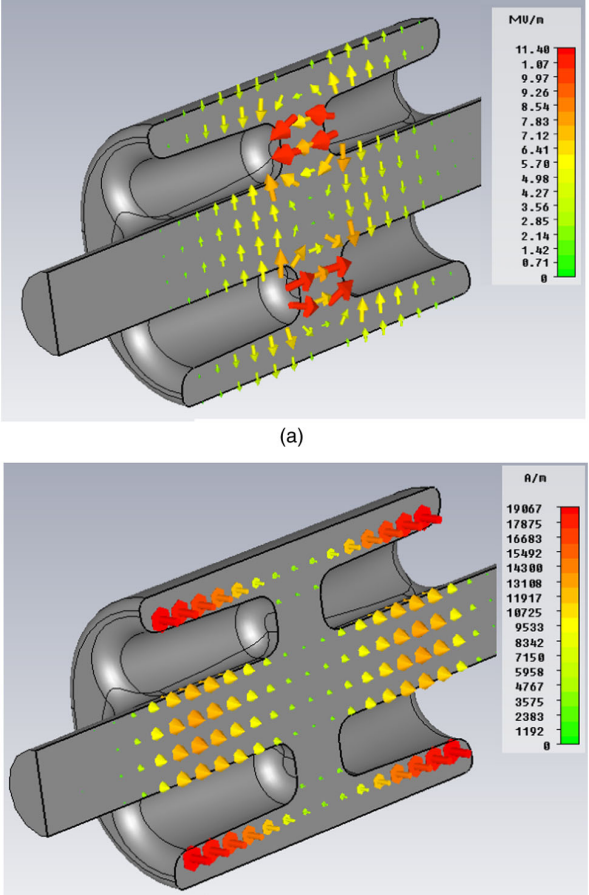}}}}}
\caption{\label{fig:4rod} The electric and magnetic fields in a 4-rod deflector \cite{bib:4rod}.}
\end{figure}

1. We use a pair of longitudinal rods that support a TEM mode between them. We have to split each rod in their middle creating a longitudinal electric field between the gap, shown in Fig. \ref{fig:4rod} \cite{bib:4rod}. Each rod should be approximately a quarter wavelength long and the dimensions of the outer can do not impact greatly on the performance so this cavity can be very compact. This type of cavity is known as a~four-rod deflecting cavity and is currently being used as a beam separator at CEBAF \cite{bib:ERLseperator}. The advantage of this type of cavity is that the TM-like modes of the cavity are at much higher frequencies compared to the~operating mode.

\begin{figure}[!h]
\centerline{\fbox{\hbox to 24pc{\vbox to 10pc{
\includegraphics[width=0.6\linewidth]{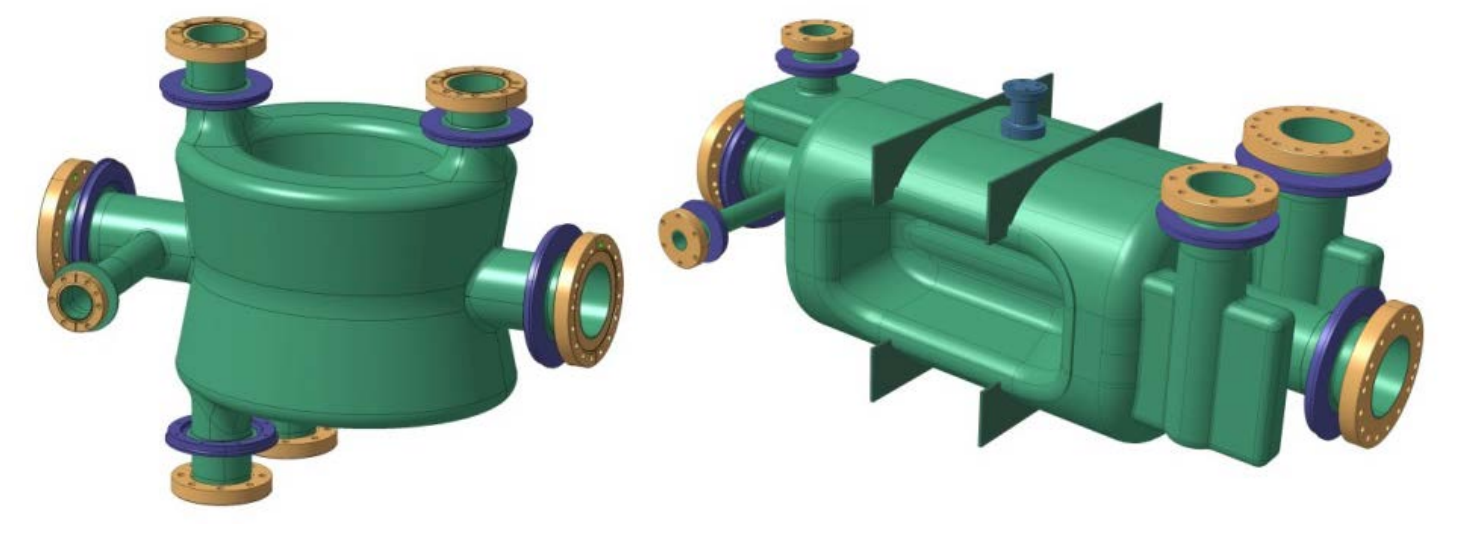}}}}
}
\caption{\label{fig:LHCcrab} The double quarter-wave (left) and RF dipole (right) crab cavities for HL-LHC \cite{bib:LHCcrab}.}
\end{figure}

2. We can inject the beam in to the cavity transversely similar to the approach used in accelerating half-wave resonators, except we need to put a capacitive gap in the middle of the inner conductor, creating two opposite quarter-wave resonators to flip the polarity across the gap to give a deflecting rather than accelerating mode \cite{bib:DQW}. This type of cavity is known as a Double Quarter-wave (DQW). Similar to this we can use a TE-mode cavity using ridged waveguide, with a gap at either end of the ridge \cite{bib:RFD}. This is known as an RF dipole (RFD). While one is TEM-like and the other is TE-like, they looks very similar as both mode are hybrid, except the DQW body is a transverse cylinder and is hence wider, and the RFD body is either rectangular or a longitudinal cylinder and is hence wider, as shown in Fig. \ref{fig:LHCcrab}. Neither cavity type has a lower order mode, making HOM damping comparatively simpler. Both of these are used in the HL-LHC \cite{bib:LHCcrab}.

\end{document}